\documentclass[%
 reprint,%
nofootinbib,
 amssymb, amsmath,%
 aip,prl%
groupedaddress,%
]{revtex4-1}

\usepackage{docs}%
\usepackage{bm}%
\usepackage[colorlinks=true,linkcolor=blue]{hyperref}%
\usepackage{graphicx}
\expandafter\ifx\csname package@font\endcsname\relax\else
 \expandafter\expandafter
 \expandafter\usepackage
 \expandafter\expandafter
 \expandafter{\csname package@font\endcsname}%
\fi
\begin{document}

\title{On extraction of value of axial mass from MiniBooNE neutrino quasi-elastic double differential cross section data}%

\begin{abstract}
MiniBooNE charge current quasi-elastic double differential cross section data are analyzed and confronted with predictions of two theoretical nucleus models: Fermi gas and spectral function. The fitting procedure includes the overall flux uncertainty multiplicative factor. In order to obtain a reliable value of the axial mass, bins with large contribution from small momentum transfer are eliminated from the analysis.  It is shown that the best fit axial mass value becomes smaller as the momentum transfer cut is more restrictive. For $q_{cut}=500$~MeV/c the obtained values of axial mass are $M_A=1350\pm 66$~MeV for the Fermi gas and $M_A=1343\pm 60$~MeV for the spectral function. The value $M_A=1030$~MeV is excluded on the level which goes far beyond $5\sigma$.
\end{abstract}

\author{Cezary Juszczak}
\author{ Jan T. Sobczyk}
\email{jsobczyk@ift.uni.wroc.pl}
\author{ Jakub $\dot{Z}$muda}%
\affiliation{Institute of Theoretical Physics\\ Wroc\l aw University}%

\date{July 13, 2010}%

\maketitle


\section{Introduction}

The knowledge of neutrino interactions in the $\sim~1$~GeV energy region is important because this energy domain is characteristic for majority of neutrino oscillations experiments performed during recent $\sim 5$ years and also those scheduled for the near future. The list includes K2K, MiniBooNE, SciBooNE, MINOS, T2K and NO$\nu$A. \\
\\
Neutrino oscillations are the energy dependent phenomenon and the most straightforward analysis of experimental data requires reconstruction of a neutrino energy. The neutrino flux spectrum is typically rather wide-band (despite significant improvement introduced with the idea of off-axis beams) and the interacting neutrino energy must be estimated based on the observation of the leptonic and/or hadronic final states. The precision of the  analysis depends on the knowledge of neutrino interaction cross-sections, both inclusive and  exclusive in several most important channels.

In this paper we discuss the charged current quasielastic reaction (CCQE)

\begin{equation}
\nu_\mu + n \rightarrow \mu^- + p.
\end{equation}
This is the dominant process in the case of sub-GeV beams in MiniBooNE, SciBooNE and T2K experiments. The theoretical description of this reaction is based on the conserved vector current (CVC) and the partially conserved axial current (PCAC) hypotheses. After a simple analysis only one  unknown quantity remains, the axial form-factor $G_A(Q^2)$, for which one typically asssumes the dipole form with an unknown parameter called the axial mass $M_A$ \cite{basic_qe}. If the deviations of $G_A(Q^2)$ from the dipole form are of similar size as those measured in the electron-nucleon scattering, it would be very difficult to observe them and the basic assumptions described above seem to be well justified. Thus the aim of CCQE cross section measurements is to estimate the value of $M_A$. Even if in a detector neutrinos interact with bound nucleons, the reported results should always refer to the parameter in the formula for free nucleon scattering. Obviously, any such measurement done on a nucleus target contains a bias from the nucleus model used in the data analysis.\\
\\
Measurements of $M_A$ typically focus on investigation of the $Q^2$ differential cross-section shape  that turns out to be sensitive enough for precise evaluation of $M_A$. Such approach has the advantage of not relying on a knowledge of the overall neutrino flux that usually carrries much uncertainty. The dependence of the total cross-section on $M_A$ could be used as a tool to fix its value provided that the overall flux is known with  high precision. The limiting value of the CCQE cross-section as $E_\nu\rightarrow\infty$ can be calculated in the analytical way assuming only dipole vector and axial form-factors \cite{ankowski}. In the exact formula the cross section dependence on $M_A$ is quadratic but in the physically relevant region with a good precision it can be considered linear. If the value of $M_A$ is increased from $1.03$ to $1.35$~GeV the cross-section and thus the expected number of CCQE events is raised by $\sim 30\%$, which is a huge effect.\\
\\

In the past there were several measurements of $M_A$,  often on the deuterium target, and until few years ago it seemed that the results converge to a value of the order of $1.03$~GeV. There is a complementary argument in favor of a similar value of $M_A$ coming from the weak pion-production at low $Q^2$. PCAC based computations give the value of $1.077\pm 0.039$~GeV \cite{MA_PCAC}. On the other hand, almost all (with the exception of the NOMAD experiment) recent high statistics measurements of $M_A$ report much larger values, see Table \ref{table_MA_results}. \\

\begin{table}[t!]
\begin{tabular}{|c|c|c|c|}
\hline
 & & & \\
 Experiment & Target & Cut in $Q^2$ [GeV$^2$]& $M_A [GeV]$ \\
 & & & \\ 
  \hline
  & & & \\
K2K\cite{k2k_oxygen_MA} & oxygen & $Q^2>0.2$ & $1.2\pm 0.12$ \\
 & & &   \\
K2K\cite{k2k_carbon_MA} & carbon & $Q^2>0.2$ & $1.14\pm 0.11$  \\
 & & &   \\
MINOS\cite{minos_MA} & iron & no cut & $1.19\pm 0.17$        \\
 & & &   \\
MINOS\cite{minos_MA} & iron & $Q^2>0.2$ & $1.26\pm 0.17$        \\
& & &  \\
MiniBooNE\cite{MB_MA} & carbon & no cut & $1.35\pm 0.17$   \\
  & & &  \\
MiniBooNE\cite{MB_MA} & carbon & $Q^2>0.25$  & $1.27\pm 0.14$  \\
 & & &  \\
\hline \hline 
& & & \\
NOMAD\cite{nomad_MA} & carbon & no cut & $1.07\pm 0.07$   \\
& & &  \\
 \hline
\end{tabular}
\caption{ Recent $M_A$ measurements 
\label{table_MA_results}}
\end{table}

There are a few possible explanations of the discrepancy. In the simplest one, this disagreement is treated as a result of statistical fluctuations (the effect is on the $< 2\sigma$ level). There are however several independent measurements and the question arises if nuclear effects may be responsible for the problem. Monte Carlo events generators used in the data analysis rely on the Impulse Approximation and the Fermi gas model and perhaps both assumptions (later in the text we explain their limitations) do not allow for a correct extraction of the value of $M_A$. \\
\\
In this paper we investigate the recently released MiniBooNE's high statistics flux averaged CCQE double differential cross section  data in the muon observables: scattering angle and kinetic energy \cite{MB_MA}. The data include corrections for detector efficiency and much effort was taken to make them independent from the nuclear physics assumptions of the MC code used in the analysis. The data provides an unprecedented possibility to validate predictions from various theoretical models.\\
\\
The MiniBooNE large values of $M_A$ were obtained from the investigation of the shape of the distribution of events in $Q^2$ and also as a fit to the normalized cross-section, and both evaluations do agree. An important element of the data analysis was a subtraction from the sample of QE-like events (no pion in the final state) of those that are believed to be not QE in the primary interaction. In the analysis NUANCE \cite{nuance} MC event generator based on the Fermi gas model was used. Obviously such subtraction depends on the ingredients of the MC model. MiniBooNE collaboration {\it corrected} the MC prediction for this background by a function that was obtained by comparing a sample of SPP-like (a single pion is detected in the final state) events to the predictions of the same MC generator. The shape of the correction function is rather poorly understood but it has an obvious and presumably important impact on the extracted value of $M_A$. The function quantifies the lack of knowledge in describing processes like pion absorption and this affects the understanding the QE-like and SPP-like samples of events. \\
\\
In order to evaluate precision of nucleus target $M_A$ measurements it is necessary to understand well all the nuclear effects. In the first place, in order to be able to get a value of the parameter of neutrino {\it free nucleon}  scattering one assumes that the reaction occurs on individual quasi-free nucleons (Impulse Approximation - IA). This is well justified if typical values of the momentum transfer are sufficiently large ($q\geq 350-400$~MeV/c, some authors speak about $q\geq 500$~MeV/c). To the contrary of what might be expected, in the case of neutrino QE interactions a fraction of at least $15\%-20\%$ of the total cross-section, independently on the neutrino energy (for neutrino energies $E_\nu<500$~MeV the situation is even worse) \cite{artur_ia}, comes from lower values of the momentum transfer. This manifests itself as the low $Q^2$ (typically $Q^2<\sim 0.1$~GeV$^2$) problem reported in several neutrino experiments: the number of events in this region is smaller than expected. This is why in the data analysis very often (see above) appropriate cuts are imposed. Since $q>\omega$ where $\omega$ is the energy transfer, and $Q^2=q^2-\omega^2$, the region of the failure of the IA is contained in the region  $Q^2< 0.1$~GeV$^2$. Experimental groups invented some ad hoc solutions to deal with the low $Q^2$ problem. The MiniBooNE collaboration proposed an effective  parameter $\kappa$ to increase the effect of Pauli blocking \cite{kappa}. CCQE fits were done simultaneously to $M_A$ and $\kappa$, treated as free parameters. In the recent MiniBooNE's paper \cite{MB_MA} the best fit to $\kappa$ is within $1\sigma$ consistent with $\kappa =1$ (no modification of the Pauli blocking). It is also important that the one-parameter fit for $M_A$ (with $\kappa =1$) does not lead to significantly different results. Also the MINOS collaboration proposed an ad hoc modification of the Pauli blocking \cite{minos_MA}. On the theoretical side, from the electron scattering data analysis it is known that the correct treatment of nucleus in the low momentum transfer region must account for collective effects (giant resonances) and computational techniques like RPA or better CRPA should be applied \cite{rpa_luis}. The impact of the limitations of the IA on the extracted value of $M_A$ will be discussed in detail in the next section. The important result of our investigation is that cuts on the momentum transfer make the fitted value of $M_A$ smaller, but the effect is by no means sufficiently strong to explain the discrepancy with the old deuterium measurements.\\
\\
FG model is determined by only two parameters: Fermi momentum $p_F$ and binding energy $B$ and it defines the probability distribution of finding inside nucleus a nucleon with a given value of momentum (quadratic distribution for $p<p_F$) and binding energy (the constant value). From the electron scattering experiments it is known that for large enough values of the momentum transfer, in the region of the {\it quasi-elastic peak} (the terminology used in the electron scattering community), the FG model allows for a reasonable agreement with the data. The advantage of the model lies in the simplicity and its MC implementation is straightforward. However, from a closer investigation of the electron scattering data, it is known that the FG model is unable to reproduce correctly the longitudinal and transverse nuclear response functions.\\
\\

\begin{figure}[t!]
\centering{
\includegraphics[width=0.5\textwidth]{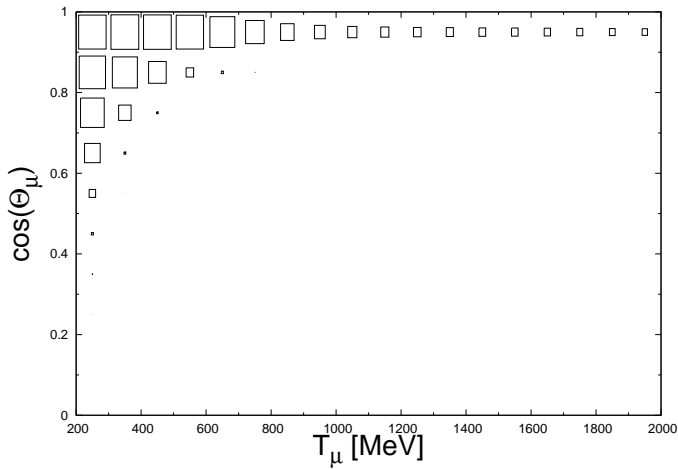}}
\caption{Contribution of events with momentum transfer lower than $q_{cut}=400$~MeV/c for the spectral function model. For each bin the contribution is proportional to the area.
\label{q_contr_400_SF}
}
\end{figure}

\begin{figure}[t!]
\centering{
\includegraphics[width=0.5\textwidth]{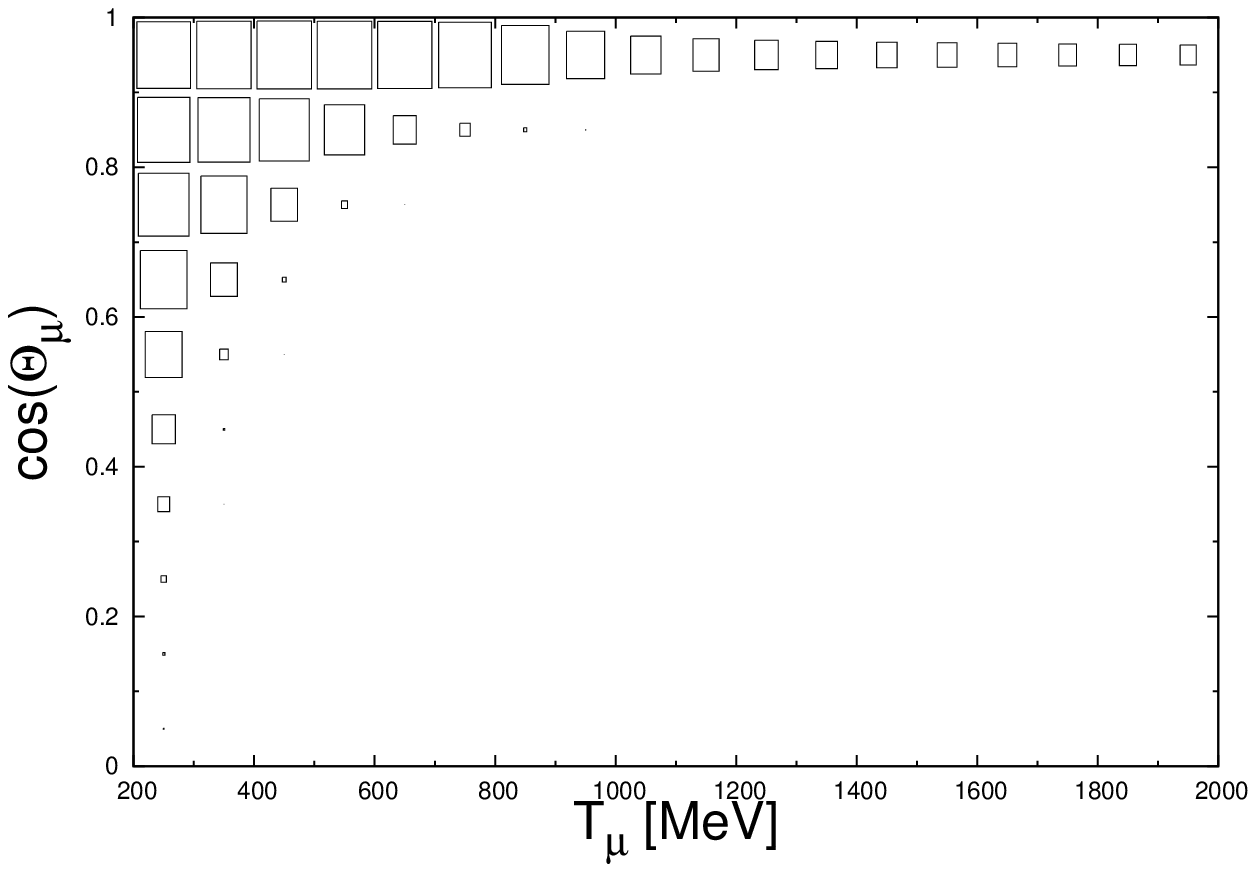}}
\caption{Contribution of events with momentum transfer lower than $q_{cut}=500$~MeV/c for the spectral function model. For each bin the contribution is proportional to the area.
\label{q_contr_500_SF}
}
\end{figure}

\begin{figure}[t!]
\centering{
\includegraphics[width=0.5\textwidth]{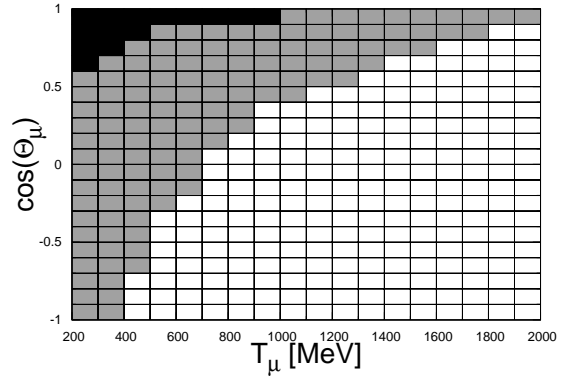}}
\caption{Bins excluded from the fitting procedure for $q_{cut}=400$~MeV/c are shown in black color. Bins with non-zero cross section measured by MiniBooNE are shown in grey color.
\label{q_excl}
}
\end{figure}

There is a variety of more sophisticated approaches to describe nuclear effects in electron scattering which were later also applied to neutrino interactions. Many of them are described in \cite{steve_nuint09} \cite{luis_review} \cite{tina_phd}. In our investigation we use the {\it spectral function} (SF) approach \cite{spectral_function}. In the context of neutrino interactions the use of SF has been advocated by Omar Benhar. SF is defined as a joint probability distribution to find inside nucleus a nucleon with a given momentum and binding energy. SF arises naturally as one calculates the neutrino QE cross-section in the plane wave impulse approximation (PWIA) \cite{pwia} i.e. assuming that the nucleon in the final state  leaves the nucleus after primary interaction with no FSI effects. The SF model leads to a very good agreement with the electron-nucleus cross-section data in the quasi-elastic region for momentum transfers larger than $\sim 350$~MeV \cite{sf_fsi} \cite{sf_electron_data}. The available models of SF combine information from the mean field theory (shell model) and a contribution from the short range correlations (SRC). The shell model orbitals are clearly seen as wide slopes in the probability distribution. SRC part contributes to the large nucleon momentum tail. In our investigation we use the implementation of the SF formalism in the NuWro MC events generator \cite{nuwro}. For carbon, oxygen and  iron NuWro uses tabularized spectral functions provided by Benhar. There also exist approximate models of SF for medium-sized nuclei like calcium and argon, which were shown to lead to a good agreement with the electron scattering data \cite{ankowski_sobczyk_2}. \\
\\
\section{Results}

\subsection{Definition of $\chi^2$}

The MiniBooNE double differential CCQE cross section data is provided in the form of the table \cite{MB_MA}. The $\cos\theta$ domain is divided into 20 equal bins, of width $0.1$. In the case of $T_\mu$ the bins have $100$~MeV width and span the region from $200$ to $2000$~MeV. Altogether they are 360 bins. The double differential cross section is non-zero in 137 of them. There is also the data for the single differential cross section in $Q^2$ in the form of 17 bins covering the region from 0 to $2$~GeV$^2$.\\
\\
Till now the fits to $M_A$ were done only on the $d\sigma/dQ^2$ data. The MiniBooNE collaboration reported the value $M_A=1.35\pm0.17$~GeV and in the recent paper Butkevich \cite{butkevich} obtained the values $1.37\pm 0.05$ and $1.36\pm0.05$ for two theoretical models used in the analysis (in the author's nomenclature: RDWIA - relativistic distorted wave impulse approximation, RFGM - relativistic Fermi gas model). The agreement is very good which is an interesting result because RDWIA is a sophisticated model which includes contribution from short range correlated nucleon pairs and corrections from FSI effects. In the fitting procedure the impact of the overall (correlated) flux uncertainty was not taken into account.\\
\\
We use the complete set of MiniBooNE'e data and for the first time make a fit to two-dimensional distribution of events in the form of double differential cross section. On the theoretical side we compare two models: Fermi gas and spectral function both implemented in the NuWro MC events generator. In the case of FG the parameters used in the simulations were: $p_F=220$~MeV/c and $B=34$~MeV. SF approach is parameter free. Pauli blocking is imposed in both models. In the case of SF the Fermi momentum value was calculated within the local density approximation. \\
\\
The samples of events were produced by NuWro for both FG and SF models for the values of the axial mass changing in steps of $10$~MeV in the $1-2$~GeV region.\\
\\
It is well known that for the same value of $M_A$ FG and SF predict quite different values of the total CCQE cross section and one could expect that the fitting procedure will give rise to very different values of $M_A$ for the two models. In the recent paper \cite{benhar_MBCCQE} the conclusion is drawn that for the SF approach the best agreement with the data is obtained with $M_A=1.6$~GeV. The total flux averaged SF CCQE cross section at $M_A=1.6$~GeV is approximately the same as the total flux averaged FG CCQE total cross section at $M_A=1.3$~GeV.\\
\\
Despite many  efforts there is still a lot of uncertainty in the knowledge of the neutrino flux \cite{kopp_flux}. MiniBooNE collaboration estimates the overall fully correlated uncertainty as 10.7\%. It is known that some other MiniBooNE measurements report larger than expected cross sections \cite{problemy_MB} which are difficult to reproduce with standard theoretical models. On the other hand the reported ratio CCPi+/CCQE is in reasonable agreement with many models \cite{mb_ratio}. It seems to be well motivated to include in the data analysis also the contribution coming from the fully correlated flux uncertainty. We apply the method of D'Agostini \cite{de_agostini} and we construct the appropriate $\chi^2$ function:

\begin{widetext}
\begin{equation}
\label{chi2} \chi^2_{} (M_A, \lambda) =\sum_{i=1}^{n_{}}
\left(\frac{\displaystyle\left( \frac{d^2\sigma } {dT_\mu d\cos\theta }\right)^{exp}_j  - \lambda \left( \frac {d^2\sigma } {dT_\mu d\cos\theta }(M_A) \right)^{th}_j }{\displaystyle  \Delta \left( \frac { d^2\sigma} { dT_\mu d\cos\theta} \right)_j } \right)^2 +
\left(\frac{\lambda^{-1} -1}{ \Delta\lambda }\right)^2.
\end{equation}
\end{widetext}
$\left( \frac{d^2\sigma } {dT_\mu d\cos\theta }\right)^{exp}_j $ is the measured double differential cross section in the j-th bin with the uncertainty 
$\Delta \left( \frac { d^2\sigma} { dT_\mu d\cos\theta} \right)_j$ (all the uncertainies are also provided by MiniBooNE). $\left( \frac {d^2\sigma } {dT_\mu d\cos\theta }\right)^{th}_j$ is the theoretical prediction from either FG or SF model for a fixed value of $M_A$. $\Delta\lambda = 0.107$ is the overall normalization uncertainty. The similar $\chi^2$ was succesfully applied in the reanalysis of the single pion production bubble chamber experiments data \cite{deuterium}. \\
\\
We investigated also the possible impact of the boundary bins in which MiniBooNE reported the vanishing cross section. For this aim we added those bins to the analysis and assumed that the uncertainty with which the null cross section is measured is equal to the average of uncertainties from all the neighbouring bins. The proposed extention of the fitting procedure allows for a {\it punishment} of the models/parameter values which give rise to too large predictions in the kinematical region excluded by the MiniBooNE measurements. Our result is that such an extention has a very small impact on the final results, shifting the best fit value of the axial mass by a few MeV only. In what follows we present the results for the $\chi^2$ calculated on the non-zero bins only.

\begin{figure}[t!]
\centering{
\includegraphics[width=0.5\textwidth]{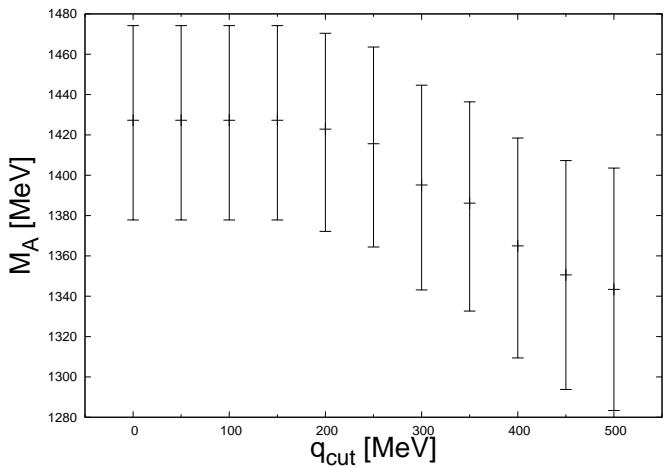}}
\caption{Best fit values of $M_A$ and $1\sigma$ regions for SF model as functions of the low momentum transfer cut.
\label{main_SF}
}
\end{figure}

\begin{figure}[t!]
\centering{
\includegraphics[width=0.5\textwidth]{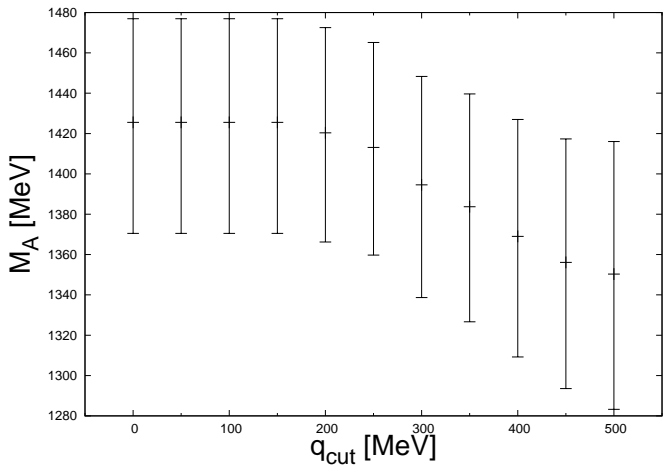}}
\caption{Best fit values of $M_A$ and $1\sigma$ regions for FG model as functions of the low momentum transfer cut.
\label{main_FG}
}
\end{figure}

\subsection{Momentum transfer cut}

We propose further refinement in the analysis. We exclude from the $\chi^2$ expression Eq. \ref{chi2} the bins with large contribution from events with small momentum transfer. The motivation was explained in the introduction: one cannot expect that the models based on the IA give reliable results in this kinematical region. In the paper \cite{rpa_luis} it was shown that inclusion of RPA correlations improves the agreement in the distribution of events in the small $Q^2$ region. In order to investigate the effect we introduce the momentum transfer cut parameter $q_{cut}$ and change its value in steps of $50$~MeV/c. The parameter is defined in such a way that the bins for which the contribution from $q<q_{cut}$ is larger than 50\% are eliminated. In the Figs \ref{q_contr_400_SF} and \ref{q_contr_500_SF} we show the contributions of events with the momentum transfer  $q<q_{cut}=400$~MeV/c and $q<q_{cut}=500$~MeV/c for the SF, in every bin separately. The results for the FG are very similar and there is no interest to show them independently. The bins which are excluded from the fitting procedure are shown in the Fig. \ref{q_excl} as marked in the black color. For every value of $q_{cut}$ the same bins survive for both FG and SF models. For the value $q_{cut}=500$~MeV/c there are still 108 bins taken into account in the numerical analysis.\\
\\
In the recent paper Butkevich\cite{butkevich} concludes that in some bins the predictions of the RDWIA and RFGM models do not agree with the data. It is an interesing observation that most of these bins are excluded from our analysis. For example at $q_{cut}=400$~MeV/c only one bin pointed out by Butkevich is present in our $q_{cut}=500$ analysis: $T_\mu\in (400, 500)$~MeV and $\cos\theta\in (0.7, 0.8)$.

\subsection{Main result}

\begin{figure}[t!]
\centering{
\includegraphics[width=0.5\textwidth]{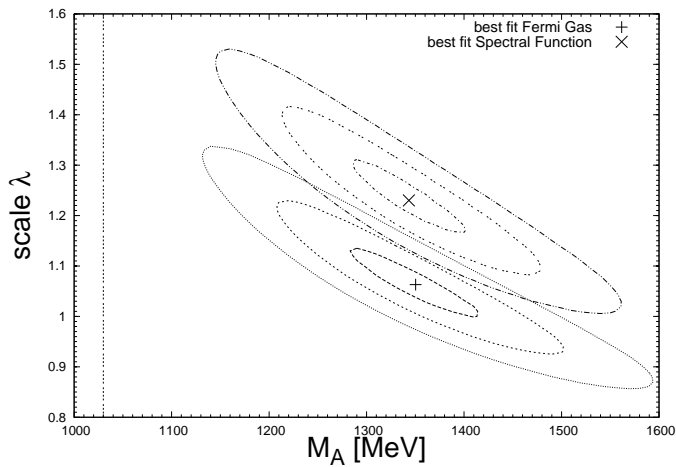}}
\caption{Best fit values of $M_A$ and $\lambda$ together with $1-\ 3-\ 5\sigma$ regions for the $q_{cut}=500$~MeV/c transfer momentum cut. We mark also with a line the old world average value of the axial mass.
\label{sigma}
}
\end{figure}

Figs \ref{main_SF} and \ref{main_FG}  contain our main discovery: the best fit values of $M_A$ for various choices of $q_{cut}$ for both SF and FG models. Contrary to what might be expected the values of the best fits for the SF model are slightly smaller than for the FG model. The reason is in the interplay between $M_A$ and $\lambda$ parameters: the best fit for $\lambda$ is in the case of SF much larger. For $q_{cut}=500$~MeV/c $\chi^2$ becomes minimal at $\lambda=1.063$ for FG and $\lambda=1.23$ for SF. The obtained best fit values for FG and SF are very similar: $M_A=1350\pm 66$~MeV for FG and $M_A=1343\pm 60$~MeV for SF. The minimal values of $\chi^2_{min}$ are different, in the case of FG they are always smaller. For example, for $q_{cut}=500$~MeV/c the minimal values are $\chi^2_{min}= 14.45$ (FG) and  $\chi^2_{min}= 23.2$ (SF). The goodness of fit is always excellent, because the uncertainties provided by MiniBooNE are very large. It is interesting to see that as $q_{cut}$ becomes larger the best fit value of $M_A$ gets smaller, and there is less tension with old bubble chamber measurements. The decline is noticeable but even if we take the maximal meaningful value of the cut, namely  $q_{cut}=500$~MeV/c, we are still far away from the old world average $M_A=1.03$~GeV. \\
\\
In the Fig. \ref{sigma} we show the two-dimensional $1-\ 3-\ $ and $5\sigma$ regions for $q_{cut}=500$~MeV/c. Because the best fit scale factors for both models are very different it is possible to show them in one figure. For the comparison we show also the old world average value of the axial mass $M_A=1.03$~GeV. Our conclusion is that old and new measurements are incompatible. This is the most important result of our investigation. Note that $5\sigma$ region is quite small because the $\chi^2$ function dependence on $M_A$ and $\lambda$ is polynomial-like rather than linear. \\
\\
We checked also the behavior of the best fits for $M_A$ for even more restrictive cuts in the momentum transfer. We discovered that for $q_{cut}>550$~MeV/c the best fit values start to increase but simultaneously also the $1\sigma$ regions start to grow. The behavior of $1\sigma$ regions is what might be expected because as $q_{cut}$ gets larger we loose more and more statistics and the predictions become less precise.\\
\\
Finally we note that the best fit value for the axial mass from our analysis is very close to the values obtained by MiniBooNE and Butkevich  from the 1-dimesional analysis of $d\sigma/dQ^2$. However, without the momentum transfer cut our results for $M_A$ would be higher. The advantage of our analysis is that we use the full information provided by the MiniBooNE collaboration and not only the $Q^2$ projection of the results.

\begin{figure}[t!]
\centering{
\includegraphics[width=0.5\textwidth]{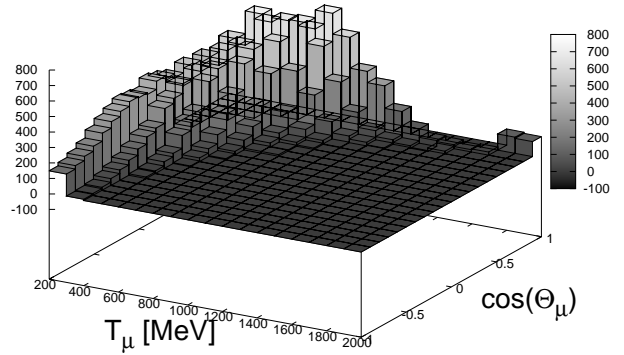}}
\caption{The difference between the double differential cross section measured by MiniBooNE and prediction from the SF model with $M_A=1.03$~GeV with no rescaling. The units are  $10^{-41}$~cm$^2$/GeV/nucleon.
\label{difference}
}
\end{figure}

\section{Conclusions}

In the comparison with the MiniBooNE CCQE double differential cross section data we used the simplest nucleus Fermi gas model which is common in MC events generators and also much more sophisticated spectral function model which is well tested on the electron scattering data in the quasi-elastic peak region. We eliminated from the discussion the low momentum transfer region where the IA based models are known to be unreliable. Our conclusion is that the new data are not compatible with the results from the old bubble chamber experiments on deuterium where the nuclear effects are easily put under control.\\
\\
It is natural to consider the possibility that the disagreement is caused by the nuclear effects which were not taken into account in the models applied so far. We know from the electron scattering that there is a need for new dynamical mechanism in the region between quasi-elastic and the $\Delta$ peaks, called the DIP region. It is known that meson exchange current (MEC) reaction in which an electron interacts with a pair of nucleons exchanging a pion adds some cross section in the DIP region, making the theoretical predictions more realistic \cite{mec}.
MEC contributes to the transverse response function where the stregth is missing. In the MEC reaction two nucleons can be ejected from the nucleus. If the analogous process would have happen in the case of $1$~GeV neutrino scattering, most likely the event would be categorized as QE-like. It is unlikely that both nucleons would be detected because they typically carry insufficient kinetic energy. Clearly such events would deceive experimentalist and contribute to the measured CCQE double differential cross section. \\
\\
There are many papers devoted to MEC in the case of electron scattering and very few evaluations of a possible significance of 2p-2h excitations in the case of neutrino scattering. According to Marteau-Martini computations \cite{marteau_martini} the contribution to the CC cross section neglected in IA models is quite large. They developed the non-relativistic model that includes QE and $\Delta$ production primary interactions, RPA correlations,  local density effects and also elementary 2p-2h excitations. The 2p-2h contribution is claimed to be able to account for the large CCQE cross-section as measured by the MiniBooNE collaboration, In the case of neutrino-carbon CCQE process after averaging over the MiniBooNE beam, nuclear effects are to increase the cross-section from 7.46 to 9.13, in the units of $10^{-39}$cm$^2$. This includes a reduction of the cross-section due to RPA effects and the increase due to the 2p-2h contribution. In the case of antineutrino-carbon CCQE reaction the RPA and 2p-2h effects cancell each other approximately, and the cross-section is virtually unchanged (modification from 2.09 to 2.07 in the same units). \\
\\
The verification of the Marteau-Martini model predictions can come from the comparison of their 2p-2h contribution with the MiniBooNE data. Because such results are not yet available, we did the comparison of the true data with the SF predictions with $M_A=1.03$~GeV and $\lambda=1$. The difference should give a contribution from the new dynamical mechanism going beyond the IA. In Fig. \ref{difference} the difference is shown with the absolute normalization. A theoretical model able to reproduce the obtained distribution would solve the MiniBooNE's axial mass puzzle.

\section*{Acknowledgements}

The authors were supported by the grant 35/N-T2K/2007/0 (the project number DWM/57/T2K/2007).

\end{document}